\documentclass[twocolumn,showpacs]{revtex4}

\usepackage{amsfonts}
\usepackage{amssymb}
\usepackage{amsbsy}
\usepackage{bm}

\begin{document}


\title{A geometrical setting for the 
classification of multilayers}

\author{Juan J. Monz\'on, Teresa Yonte, Luis L. S\'anchez-Soto}
\affiliation{Departamento de \'Optica, 
Facultad de Ciencias F\'{\i}sicas, 
Universidad Complutense, 
28040 Madrid,  Spain}

\author{Jos\'e F. Cari\~{n}ena}
\affiliation{Departamento de F\'{\i}sica Te\'orica, 
Facultad de Ciencias, Universidad de Zaragoza, 
50009 Zaragoza,  Spain}

\begin{abstract}
We elaborate on  the consequences of the 
factorization of the transfer matrix of any 
lossless multilayer in terms of three basic 
matrices of simple interpretation. By considering
the bilinear transformation that this transfer matrix
induces in the complex plane, we introduce the
concept of multilayer transfer function and study
its properties in the unit disk. In this geometrical
setting, our factorization translates into three actions 
that can be viewed as the basic pieces 
for understanding the multilayer behavior.
Additionally, we introduce a simple trace criterion
that allows us to classify multilayers in three types
with properties closely related to one (and only one)
of these three basic matrices. We apply this 
approach to analyze some practical examples that are 
typical  representatives of these types of matrices.
\end{abstract}


\maketitle

\section{Introduction}

Layered media play an important role in many applications 
in modern optics, especially in relation to optical filters 
and the like. Therefore, it is not surprising that the topics 
covered in most of the textbooks on the subject use a 
mixture of design,  manufacture,  and applications; 
dealing only with the basic physics needed  to carry out  
practical computations~\cite{MA86}.

However, for a variety of reasons, layered media have 
physical relevance on their own~\cite{YE88,LE87}. 
As any linear system with two input and two output 
channels, any multilayer can be described in terms of 
a $2 \times 2$ transfer matrix.  In fact, it has been 
recently established that for a lossless multilayer this 
transfer matrix is an element of the group 
SU(1,1)~\cite{MO99a,MO99b}. From this perspective,  
it is precisely  the abstract composition law of 
SU(1,1) the  ultimate responsible for the curious 
composition law of the reflection and transmission 
coefficients~\cite{MO99c,MO01a}.

This purely algebraic result is certainly remarkable.
But, as soon as one realizes that SU(1,1) is also the
basic group of the hyperbolic geometry~\cite{CO68}, 
it is tempting to look for an enriching geometrical
interpretation of the multilayer action. Moreover, given 
the role played by  geometrical ideas in all branches of 
physics,  particularly in special relativity,  it is easy to 
convince oneself that this approach might provide deeper 
insights into the behavior of a multilayer in a wider 
unifying framework that can put forward fruitful analogies 
with other physical phenomena.

Accordingly, we have proposed~\cite{MO02} to view 
the action of  any lossless multilayer as a bilinear 
transformation on the unit disk, obtained by stereographic 
projection of the unit hyperboloid of SU(1,1). This kind of bilinear 
representations have been discussed in detail for the Poincar\'e 
sphere in  polarization optics~\cite{AZ87,HA96}, for Gaussian 
beam propagation~\cite{KO65}, and are also useful in laser 
mode-locking and optical pulse transmission~\cite{NA98}.

In spite of these achievements, the action of an 
arbitrary lossless stack could still become cumbersome 
to interpret in physical terms. In fact, in practice it is usual 
to work directly with  the numerical values of a matrix 
obtained from the experiment, which cannot be directly 
related to the inner multilayer structure. To remedy this 
situation, we have resorted recently~\cite{MO01b}
to the Iwasawa decomposition, which provides a remarkable
factorization of the matrix representing any multilayer 
(no matter how complicated it could be)  as  the product  
of three matrices of simple interpretation.

At the geometrical level, such a decomposition translates
directly into the classification of three basic actions in the
unit disk, which are studied in this paper, that can be 
considered as the basic bricks for understanding multilayers. 
Moreover, we have also shown~\cite{MO01c} that 
the trace of the transfer matrix allows for a classification 
of multilayers in three different types with properties very 
close to those appearing in the Iwasawa decomposition.

In this paper we go one step further and exploit this 
new classification to study several practical examples that
are representatives of each type. This shows the power 
of the method and, at the same time, allows for a deeper 
understanding of layered media. As a direct application,
we treat the outstanding case of symmetric multilayers,
finding a precise criterion for zero-reflectance conditions.
Nevertheless, we stress that the benefit of this  formulation 
lies not in any inherent advantage  in terms of  efficiency 
in solving problems in layered structures.  Rather,  we  
expect that the formalism presented here could 
provide  a general and unifying tool to analyze multilayer 
performance in an elegant and concise way that, 
additionally, is closely related to other fields of 
physics, which seems to be more than a curiosity.

\section{Transfer matrix for a lossless multilayer}

We first briefly summarize the essential ingredients 
of multilayer optics we shall need for our purposes~\cite{AZ87}.
The  configuration is a stratified structure, illustrated 
in Fig.~1, that consists of a stack of  $1, \ldots, j, \ldots, m$,  
plane--parallel lossless layers sandwiched 
between two semi-infinite ambient ($a$) and substrate 
($s$) media, which we shall assume to be identical, since
this is the common experimental case. Hereafter
all the media are supposed to be lossless,  
linear, homogeneous, and isotropic. 

We consider an incident monochromatic linearly 
polarized plane wave from the ambient, which makes 
an angle  $\theta_0$ with  the normal to the first 
interface and  has amplitude  $E_{a}^{(+)}$. 
The electric field is either in the plane of  incidence 
($p$ polarization) or perpendicular to the plane of 
incidence ($s$ polarization). We consider as well 
another plane wave of the same frequency and  
polarization, and with amplitude $E_{s}^{(-)}$, 
incident from the substrate at the same  angle 
$\theta _{0}$.~\cite{Snell} 

As a result of multiple reflections in all the interfaces,
we have a backward-traveling plane wave in the 
ambient, denoted $E_{a}^{(-)}$, and a  
forward-traveling plane wave in the substrate, 
denoted $E_{s}^{(+)}$. If we take  the field  
amplitudes as a vector of the form
\begin{equation}
\label{Evec}
\mathbf{E} = 
\left ( \begin{array}{c}
E^{(+)} \\ 
E^{(-)} 
\end{array}
\right )\ ,
\end{equation}
which applies to both ambient and substrate
media, then the amplitudes at each side 
of the multilayer are related by the $2 \times 2$ 
complex matrix $\mathsf{M}_{as}$, we shall 
call the multilayer transfer matrix,~ \cite{OH00} 
in the form 
\begin{equation}
\label{M1}
\mathbf{E}_a =  
\mathsf{M}_{as} \,
\mathbf{E}_s\ .
\end{equation}
The matrix  $\mathsf{M}_{as}$ can be shown to 
be of the form~\cite{MO99c} 
\begin{equation}
\label{Mlossless}
\mathsf{M}_{as} =
\left [
\begin{array}{cc}
1/T_{as} & R _{as}^\ast/T_{as}^\ast \\ 
R_{as}/T_{as} & 1/T_{as}^\ast
\end{array}
\right ]  
\equiv
\left [
\begin{array}{cc}
\alpha & \beta \\ 
 \beta^\ast & \alpha^\ast
\end{array}
\right ]  ,
\end{equation}
where the complex numbers 
\begin{equation}
R_{as}  =  | R_{as} | \exp (i \rho), 
\qquad  
T_{as}  =  | T_{as} | \exp (i \tau) ,
\end{equation} 
are, respectively,  the overall reflection and transmission 
coefficients for a wave incident from the ambient.
Note that  we have $ \det \mathsf{M}_{as}= +1$,
which is equivalent to $|R_{as}|^2  + |T_{as}|^2 = 1$, 
and then  the set of lossless multilayer matrices
reduces to the group SU(1,1), whose elements depend
on three independent real parameters.

The identity matrix corresponds to $T_{as} =1$ and 
$R_{as} = 0$, so it  represents an antireflection system
without transmission phase shift.  The matrix that 
describe the overall system obtained by  putting two 
multilayers together is the product of the matrices 
representing each  one of them, taken in the 
appropriate order.  So, two  multilayers, which are inverse, 
when composed give an antireflection system. 

If  we denote by $R_{sa}$ and $T_{sa}$ the overall 
reflection and  transmission coefficients for a  wave 
incident  from the substrate (physically this corresponds 
to the same multilayer taken in the reverse order) one 
can check that~\cite{MO99a}
\begin{eqnarray}
\label{Stokes}
& T_{as}T_{sa} -  
R_{as}R_{sa}  =   \exp (i 2\tau), &
\nonumber \\
& & \\
& R_{sa} =  - R_{as}^\ast  \exp (i 2 \tau) , &
\nonumber
\end{eqnarray}
which is a generalization of the well-known 
Stokes relations~\cite{LE87} for the overall
stack. 

On the other hand, it is worth noting that,  while the 
picture of a multilayer taken in  the reverse order is clear, 
at first sight it is not so easy to imagine the inverse of that 
multilayer. However, using Eqs.~(\ref{Stokes}), one can 
obtain that 
\begin{equation}
\mathsf{M}^{-1}_{as}= 
\mathsf{M}^{\ast}_{sa} ,
\end{equation} 
which remedies this drawback.

\section{A basic factorization for multilayers: 
the Iwasawa decomposition}
 
Many matrix factorizations have been considered 
in the literature~\cite{AR83, AB94,SH95}, the goal
of all of them being to decompose a matrix as a unique 
product  of other matrices of simpler interpretation. 
Particularly, given the essential role  played by  the 
Iwasawa decomposition, both in fundamental 
studies and in applications  to several fields (especially 
in  optics), one is tempted to investigate also its role in 
multilayer optics.  

Without embarking us in mathematical subtleties,  the  
Iwasawa decomposition is established  as follows~\cite{HE78}: 
any element of a (noncompact semi-simple) Lie group 
can be written  as an ordered product of three elements, 
taken one each from a maximal compact subgroup 
$\mathsf{K}$, a maximal Abelian subgroup $\mathsf{A}$, 
and  a maximal nilpotent subgroup $\mathsf{N}$. Furthermore, 
such a  decomposition  is global and unique.

For a lossless multilayer matrix $\mathsf{M}_{as}  
\in$  SU(1,1),  the decomposition reads 
as \cite{MO01b}
\begin{equation}
\label{Iwa1}
\mathsf{M}_{as} = \mathsf{K}(\phi)\,  \mathsf{A}(\xi)\, 
\mathsf{N}(\nu)\ ,
\end{equation}
where
\begin{eqnarray}
\label{Iwasa1}
\mathsf{K}(\phi) & = & 
\left [
\begin{array}{cc}
\exp (i\phi/2) & 0 \\ 
0 & \exp (-i\phi/2)
\end{array}
\right ]\  , 
\nonumber \\
\mathsf{A}(\xi) & = & 
\left [
\begin{array}{cc}
\cosh (\xi/2) & i\, \sinh(\xi/2) \\ 
-i\, \sinh(\xi/2) & \cosh (\xi/2)
\end{array}
\right ]\ , \\
\mathsf{N}(\nu) & = & 
 \left [
\begin{array}{cc}
1 - i\, \nu/2& \nu/2 \\ 
\nu/2 & 1+ i\, \nu/2
\end{array}
\right ]\  .
\nonumber
\end{eqnarray}
The parameters $\phi, \xi$,
and $\nu$ are given in terms of the
elements of the multilayer matrix by
\begin{eqnarray}
\label{param}
\phi/2 & = & \arg (\alpha + i \beta)\ ,  \nonumber  \\
\xi/2  &  =  & \ln  (1/|\alpha +i \beta | )\ , \\
\nu/2 & = & \mathrm{Re} (\alpha  \beta^\ast)/
|\alpha + i \beta |^2\ ,  \nonumber 
\end{eqnarray}
and their ranges  are  $\xi,  \nu \in \mathbb{R}$ 
and $-2\pi \le \phi \le 2\pi$. Therefore, given 
\textit{a priori} limits on $\alpha$ and 
$\beta$ (i.e., on $T_{as}$ and $R_{as}$), one could
easily establish the corresponding limits on 
$\phi$, $\xi$ and $\nu$, and vice versa.

Now, we can interpret the physical action of the 
matrices appearing in Eq.~(\ref{Iwa1}).  
${\mathsf{K}}(\phi)$  represents the free propagation 
of the fields  $\mathbf{E}$ in the ambient medium through
an optical phase thickness of $\phi/2$, which  reduces 
to a mere shift of the origin of phases.  Alternatively,
this can be seen as an antireflection system. The 
second matrix ${\mathsf{A}}(\xi)$  represents  a 
symmetric multilayer with real transmission 
coefficient  $T_{\mathsf{A}} = \mathrm{sech}(\xi/2)$ 
and  reflection phase shift  $\rho_{\mathsf{A}} = \pm \pi/2.$ 
There are many ways to get this performance, 
perhaps the simplest one is a Fabry-Perot system 
composed by two identical plates separated by a 
transparent spacer. By adjusting the  refractive indices 
and the thicknesses of the media one can always get 
the desired values (see Section 7). Finally, the third 
matrix, ${\mathsf{N}}(\nu)$, represents a system 
having $T_{\mathsf{N}} = \cos(\tau_{\mathsf{N}}) 
\exp(i \tau_{\mathsf{N}})$  and $R_{\mathsf{N}} =  
\sin(\tau_{\mathsf{N}}) \exp(i \tau_{\mathsf{N}})$,  with  
$ \tan (\tau_{\mathsf{N}}) =  \nu/2$. The simplest way 
to accomplish this task is by an asymmetrical 
two-layer system. 

\section{A remarkable case: symmetric multilayers}

So far, we have discussed various properties of
arbitrary lossless multilayers. However, the
particular case of symmetric structures has 
deserved a lot of attention. In this Section, we wish 
to show how the Iwasawa decomposition provides a 
nice tool to deal with them.

We recall that for a symmetric stack the reflection 
and transmission  coefficients are the same whether 
light is incident on  one side or on the opposite side of 
the  multilayer. In consequence, one has   $R_{as} = 
R_{sa}$  and the generalized Stokes relations 
(\ref{Stokes}) give then the well-known 
result~\cite{GI80,ZE81,OU89}
\begin{equation}
\rho - \tau = \pm \pi/2 .
\end{equation}
This implies that the element $\beta$ in the transfer 
matrix  (\ref{Mlossless}) is a pure imaginary number.
Therefore, the matrix $\mathsf{M}_{as}$ depends 
only on two real parameters, which translates into the 
fact that $\phi$, $\xi$, and $\nu$ [see Eqs.~(\ref{param})] 
are not independent. In fact, a straightforward calculation 
shows that they must fulfill the constraint
\begin{equation}
\label{res}
\nu = (e^\xi - 1) \tan(\phi/2) ;
\end{equation}
and thus $R_{as}$ can be written as
\begin{equation}
\label{Rsym}
R_{as} = e^{- i \phi} \frac{\tanh(\xi/2)
[ \tan (\phi/2) - i ]}{1 - i  \tanh(\xi/2)
\tan(\phi/2)} .
\end{equation}

Particular care has been paid to the characterization
of zero reflectance conditions for these symmetrical 
systems~\cite{LE89,LE90}.  Equation~(\ref{Rsym})
allows us to express the locii of zero $R_{as}$ by
the simple condition
\begin{equation}
\label{xi}
\xi = 0 ,
\end{equation}
which, by Eq.~(\ref{res}) implies $\nu = 0$ and
imposes that in this case $\mathsf{M}_{as}$ 
reduces trivially to a matrix $\mathsf{K}(\phi)$.

The stability of these nonreflecting configurations 
has been studied by Lekner~\cite{LE90}. Indeed, by
using a continuity argument, he has shown that
``almost all partial reflectors with symmetric 
profiles which are close in parameter space to a profile
which has reflectivity zeros, will also have reflectivity
zeros". We intend to show how our formalism allows
for a more precise criterion.

To this end, let us assume a symmetric multilayer 
satisfying initially the condition (\ref{xi}). Now,
suppose that some parameter(s) (refractive index, 
thickness, angle of incidence, ...), we shall 
generically denote by $ \ell$, is varied. Obviously,
admissible variations must preserve the symmetry
of the system. 

The variation of $\ell$ induces changes in $R_{as}$, 
and so in $\phi$ and $\xi$.  The new multilayer will 
have also zero $R_{as}$ if  the  parameters satisfy 
$d R_{as}/d\ell = 0$; that is,
\begin{equation}
\frac{dR_{as}}{d \ell} =
\left . \frac{\partial R_{as}}{\partial \phi} 
\right |_{\xi = 0} \frac{d \phi}{d \ell}  +
\left . \frac{\partial R_{as}}{\partial \xi} 
\right |_{\xi = 0} \frac{d \xi}{d \ell}  = 0 .
\end{equation}
Using Eq.~(\ref{Rsym}), one gets that
$\partial R_{as}/\partial \phi |_{\xi = 0} $ is identically
zero, while $\partial R_{as}/\partial \xi |_{\xi = 0} $
never vanishes. We conclude then that the
condition we are looking for is
\begin{equation}
\frac{d \xi}{d \ell}  = 0 .
\end{equation}
This result fully characterizes the partial reflectors invoked
by Lekner, and can be of practical importance for the 
design of robust antireflection systems.

\section{Multilayer transfer function in the unit disk}

In many instances (e.g., in polarization 
optics~\cite{AZ87}), we are interested in the 
transformation properties of quotients of variables 
rather than on the variables themselves. 
In consequence, it seems natural to consider the 
complex numbers
\begin{equation}
\label{defz}
z   =  \frac {E^{(-)}}{E^{(+)}} ,
\end{equation}
for both ambient and substrate. From a geometrical 
viewpoint, Eq.~(\ref{M1}) defines a transformation 
of  the complex plane ${\mathbb{C}}$, mapping the 
point $z_s$ into the point $z_a$, according to
\begin{equation}
\label{accion}
z_a = \Phi [\mathsf{M}_{as} , z_s] = 
\frac{\beta^\ast +\alpha^\ast z_s} 
{\alpha + \beta z_s}  .
\end{equation}
Thus, the action of the multilayer can be seen  as 
a function  $z_a = f(z_s)$ that can be appropriately 
called the multilayer transfer function~\cite{MO02}.
The action of the inverse matrix $\mathsf{M}_{as}^{-1}$ is 
$z_s = \Phi [\mathsf{M}_{as}^{-1}, z _a]$.

These bilinear transformations define an action of the 
group SU(1,1) on the complex plane $\mathbb{C}$. The 
complex  plane appears then decomposed  in three regions 
that remain invariant  under the action of the group: the unit disk, 
its boundary and the external region.~\cite{PE86}

The Iwasawa decomposition has an immediate 
translation in this geometrical framework, and one 
is led to treat separately the action of each one of 
the matrices appearing in this decomposition. To
this end, it is worth noting that the group SU(1,1) 
that we are considering appears always as a group of 
transformations of the complex plane. The concept 
of orbit is especially appropriate for obtaining an intuitive 
picture of the corresponding action. We recall that, given 
a point $z$,  its orbit is the set of  points $z^\prime$
obtained from $z$ by the action of  all the elements of 
the group. In Fig.~2 we have plotted some typical orbits 
for each one of the subgroups of  matrices 
$\mathsf{K}(\phi)$, $\mathsf{A} (\xi)$, and 
$\mathsf{N} (\nu)$. For matrices $\mathsf{K}(\phi)$ 
the orbits are circumferences centered at the origin and
passing by $z$. For $\mathsf{A} (\xi)$, they are arcs of 
circumference going from the point $ +i$ to the point $-i$ 
through $z$.  Finally, for the matrices $\mathsf{N} (\nu)$ the 
orbits  are  circumferences passing through the point $+ i$ and 
joining the points $z$ and $-z^\ast$.

\section{Trace criterion for classification of multilayers}

To go beyond this geometrical picture of multilayers, let us 
introduce the following classification: a matrix is of type 
$K$ when $[\mathrm{Tr}( \mathsf{M}_{as})]^2 < 4$,
is of type $A$ when  $[\mathrm{Tr} ( \mathsf{M}_{as})]^2 > 4$, 
and finally is of  type $N$ when $[\mathrm{Tr}( \mathsf{M}_{as})]^2 
= 4$. To gain insight into this classification, let us also 
introduce the fixed points~\cite{BA47} of a transfer matrix as the 
points in the complex plane that are invariant under 
the action of $\mathsf{M}_{as}$; i.e.,
\begin{equation}
\label{C}
z = \Phi[\mathsf{M}_{as}, z ] ,
\end{equation} 
whose solutions are 
\begin{equation}
z = \frac{-i \mathrm{Im}(\alpha) \pm
\sqrt{[\mathrm{Re} (\alpha)]^2 -1}}
{\beta} \ .
\end{equation}
Since we have
\begin{equation}
\mathrm{Tr} (\mathsf{M}_{as}) =
2  \mathrm{Re}(\alpha) = \frac{2 \cos \tau}
{|T_{as}|} , 
\end{equation}
one can easily check that the matrices of type 
$K$ have two fixed points, one inside and other 
outside the unit disk, both related by an inversion; 
the matrices of type $A$ have two fixed points 
both on the boundary of the unit disk and, finally,  
the matrices of type $N$ have only one (double) 
fixed point on the boundary of the unit disk.

Now the origin of the notation for these types of matrices
should be clear: if one consider the Iwasawa 
decomposition~(\ref{Iwa1}),  one can see 
that the matrices  $\mathsf{K} (\phi)$  are of  
type $K$ with the origin as the fixed point in the unit disk,  
matrices $\mathsf{A} (\xi)$ are of type $A$ with fixed points 
$+ i$ and $-i$ and   matrices $\mathsf{N} (\nu)$ are of type 
$N$ with  the double fixed point $+ i$. Of course, this is in
agreement with the orbits shown in Fig.~2.

To proceed further let us note that by taking the conjugate
of $\mathsf{M}_{as}$ with any matrix 
$\mathsf{C}\in $ SU(1,1) we obtain another
multilayer matrix; i.e.,
\begin{equation}
\label{conjC}
\widehat{\mathsf{M}}_{as} = \mathsf{C}  \
\mathsf{M}_{as} \ {\mathsf{C}}^{-1} ,
\end{equation}
such that $\mathrm{Tr} (\widehat{\mathsf{M}}_{as}) =
\mathrm{Tr} (\mathsf{M}_{as})$. The fixed points
of $\widehat{\mathsf{M}}_{as}$ are then the image
by $\mathsf{C}$ of the fixed points of $\mathsf{M}_{as}$.
If we write the matrix $\mathsf{C}$ as
\begin{equation}
\mathsf{C} =  
\left [
\begin{array}{cc}
\ \mathfrak{c}_1  \ & \ \mathfrak{c}_2 \ \\ 
\ \mathfrak{c}^\ast_2 \  & \ \mathfrak{c}^\ast_1 \
\end{array}
\right ]\  , 
\end{equation}
the matrix elements of $\widehat{\mathsf{M}}_{as}$
(denoted by carets) and those of $\mathsf{M}_{as}$
are related by
\begin{eqnarray}
\widehat{\alpha} & = &  \alpha |\mathfrak{c}_1|^2 -
\alpha^\ast |\mathfrak{c}_2|^2 - 
2 i \mathrm{Im} (\beta \mathfrak{c}_1 
\mathfrak{c}^\ast_2 ) , \nonumber \\
& & \\
\widehat{\beta} & = &  \beta \mathfrak{c}_1^2 -
\beta^\ast \mathfrak{c}_2^2 - 
2 i  \mathfrak{c}_1 \mathfrak{c}_2 
\mathrm{Im} (\alpha) . \nonumber
\end{eqnarray}
 
For our classification viewpoint it is essential to remark 
that if a multilayer has a transfer matrix of type 
$K$, $A$, or $N$, one can always find  a family 
of matrices $\mathsf{C}$ such that $\widehat{\mathsf{M}}_{as}$
in Eq.~(\ref{conjC}) is just a matrix  $\mathsf{K}(\phi)$, 
$\mathsf{A}(\xi)$, or $\mathsf{N}(\nu)$,  respectively. The 
explicit  construction of this family of matrices is easy: 
it suffices to impose that $\mathsf{C}$ transforms 
the fixed points of $\mathsf{M}_{as}$ into the 
corresponding fixed points of $\mathsf{K}(\phi)$, 
$\mathsf{A}(\xi)$, or  $\mathsf{N}(\nu)$. By way of
example,  let us consider the case when $\mathsf{M}_{as}$ 
is of type $K$ and its fixed point inside the unit disk is 
$z_f$. Then,  one should have
\begin{equation}
\Phi[\mathsf{C} \mathsf{M}_{as} {\mathsf{C}}^{-1}, 0]
= \Phi[\mathsf{C} \mathsf{M}_{as}, z_f] = 
\Phi[\mathsf{C}, z_f] = 0 .
\end{equation}
Solving this equation one gets directly
\begin{equation}
\mathfrak{c}_1  =  \frac{1}{\sqrt{1 - |z_f|^2}}
\exp(i \delta) , \qquad 
\mathfrak{c}_2  =   - \mathfrak{c}_1 z^\ast_f ,
\end{equation}
where $\delta $ is a real free parameter. The same procedure
applies to the other two cases.

Since the matrix $\widehat{\mathsf{M}}_{as}$
belongs to one of the subgroups $\mathsf{K}(\phi)$,
$\mathsf{A}(\xi)$, or $\mathsf{N}(\nu) $ of the Iwasawa
decomposition, and all these subgroups are, in our 
special case, Abelian and uniparametric,
we have that 
\begin{equation}  
\widehat{\mathsf{M}}_{as} (\mu_1)
\widehat{\mathsf{M}}_{as} (\mu_2) 
= \widehat{\mathsf{M}}_{as} (\mu_1 + \mu_2) ,
\end{equation}
where $\mu$ represent the adequate parameter
$\phi$, $\xi$, or $\nu$. Therefore, when dealing
with a periodic layered system, whose matrix is 
obtained as the $N$th power of the basic
period, we  have 
\begin{equation}  
\mathsf{M}^N_{as} =
{\mathsf{C}}^{-1}   \widehat{\mathsf{M}}^N _{as} (\mu)
\mathsf{C} =
{\mathsf{C}}^{-1}   \widehat{\mathsf{M}}_{as} (N \mu) 
\mathsf{C} ,
\end{equation}
where $\mathsf{M}_{as}$ is now the matrix of the 
basic period. This is a remarkable result~\cite{BO99}, and 
our procedure highlights that it does not depend on the explicit 
form of the basic period.

We wish to point out that the trace criterion has been 
previously introduced~\cite{LE94,LE00} to treat the
light propagation  in periodic structures. In fact,
in that approach the values of the trace separate the
band-stop from the band-pass regions of a period stratification.

Finally, in order to broaden the physical picture of this 
classification, let us transform Eq.~(\ref{M1}) 
by  the  unitary matrix 
\begin{equation}
{\mathcal{U}} =
\frac{1}{\sqrt{2}}
\left [
\begin{array}{cc}
1 & i \\ 
i & 1
\end{array}
\right ]   .
\end{equation}
Then,  we can rewrite it alternatively as
\begin{equation}
\bm{{\mathcal{E}}}_a =  
{\mathcal{M}}_{as} 
\bm{{\mathcal{E}}}_s ,
\end{equation}
where the new field vectors $\bm{{\mathcal{E}}}$
and the new multilayer matrix $\mathcal{M}_{as}$
are obtained by conjugation by $\mathcal{U}$.

One can easily check that $\det{\mathcal{M}}_{as} = +1$ 
and all its elements are real numbers.  Therefore, 
${\mathcal{M}}_{as}$ belongs to the group SL(2,$\mathbb{R}$)
that underlies the structure of the celebrated $ABCD$ law
in first-order optics~\cite{geom1,geom2,geom3,geom4}. 

By transforming by ${\mathcal{U}}$ the Iwasawa 
decomposition (\ref{Iwa1}), we get  the corresponding 
one for SL(2,$\mathbb{R}$), which has been 
previously worked out~\cite{Iwasl2r}:
\begin{equation}
\label{Iwa2}
{\mathcal{M}}_{as} = 
{\mathcal{K}}(\phi)  
{\mathcal{A}}(\xi) 
{\mathcal{N}}(\nu) ,
\end{equation}
where
\begin{eqnarray}
{\mathcal{K}}(\phi)  & = & 
\left [
\begin{array}{cc}
\cos(\phi/2) &  \sin(\phi/2) \\ 
- \sin(\phi/2)  & \cos(\phi/2) 
\end{array}
\right ]  , 
\nonumber \\
{\mathcal{A}}(\xi) & = & 
\left [
\begin{array}{cc}
\exp(\xi/2) & 0 \\ 
0 & \exp (-\xi/2)
\end{array}
\right ] , \\
{\mathcal{N}}(\nu) & = & 
 \left [
\begin{array}{cc}
1 & 0 \\ 
 \nu & 1
\end{array}
\right ]  .
\nonumber
\end{eqnarray}
The physical action of these matrices is clear.
Let us consider all of them as $ABCD$ matrices 
in geometrical optics that apply to position 
$\mathbf{x}$ and momentum $\mathbf{p}$ 
(direction) coordinates of a ray in a transverse plane.
These are the natural phase-space variables
of ray optics. Then $\mathcal{K} (\phi)$ would 
represent a rotation in  these variables,  $\mathcal{A} (\xi)$ 
a magnifier that scales $\mathbf{x}$ up to the factor 
$m = \exp(\xi/2)$ and $\mathbf{p}$ down by the same 
factor, and $\mathcal{N} (\nu)$  the action of a lens 
of power $\nu$~\cite{geom3}.   

In the multilayer picture, ${\mathcal{E}}^{(+)}$ 
can be seen as the corresponding  $\mathbf{x}$,
while ${\mathcal{E}}^{(-)}$ can be seen as
the corresponding $\mathbf{p}$. Then, the key
result  of this discussion is that when the multilayer transfer 
matrix has  $[\mathrm{Tr} ({\mathsf{M}}_{as}) ]^2$ 
lesser than, greater than or equal to 4 one can find in a direct 
way  a family of matrices that gives a new vector basis such 
that the action of the multilayer, when viewed in such a 
basis, is exclusively rotationlike, or magnifierlike, or 
lenslike~\cite{MO01c}.

\section{Simple examples and concluding remarks}

It seems pertinent to conclude by showing how our
approach works in some practical examples. Perhaps, 
the best way of starting is to consider the simplest
layered structure one can imagine: a single film 
sandwiched between the same ambient and substrate 
media. In spite of its simplicity, it contains the essential 
physical ingredients of multilayer optics.

In consequence, we consider a single transparent
film (medium 1)  of refractive index $n_1$  and thickness 
$d_1$ embedded in air (medium 0).  For this system we 
have~\cite{AZ87}
\begin{equation}
T_{as}  \equiv T_{010} =
\frac{(1 - r_{01}^2) \exp(- i \beta_1)}
{1- r_{01}^2  \exp(- i 2 \beta_1) } , 
\end{equation}
where $r_{01}$ is the Fresnel reflection coefficient
at the interface 01 and $\beta_1= (2 \pi  n_1 d_1 
\cos \theta_1)/ \lambda$ is the plate phase thickness
(here $\lambda$ is the wavelength \textit{in vacuo} 
of the incident  light and $\theta_1$ is the refraction angle).
Accordingly, we get
\begin{equation}
[\mathrm{Tr}(\mathsf{M}_{010})]^2  = 
4 \cos^2 \beta_1 \le 4,
\end{equation}
and the equality holds only in resonance conditions;
i.e., when $| T_{010}  | = | \cos \beta_1| = 1$. In
consequence, the matrix of a single film is  always of
type $K$. 

Let us consider now two films (1 and 2) described by 
the matrices $\mathsf{M}_{010}$ and 
$\mathsf{M}_{020}$,  respectively. The 
compound system obtained by putting them 
together is described by the product of these
matrices $\mathsf{M}_{010} 
\mathsf{M}_{020}$~\cite{MO99a,MO99c,AZ87}.
In conclusion, since any layered stack can be
viewed as the composition of single films, this
shows that any multilayer matrix is generated by 
the product of matrices of type $K$. Take into
account that the product of two matrices of
type $K$ (or $A$ or $N$) can have trace lesser
than, greater than or equal to 4.

On the other hand, as we have stated at the end of Section~3, 
to get a pure matrix $\mathsf{A}(\xi)$ one should consider 
a Fabry-Perot--like system formed by two identical 
plates (each one of them with phase thickness $\beta_1$) 
separated by a spacer of air with phase thickness $\beta_2$.  
If we take as initial condition that in the substrate $z_s = 0.4 \exp(- i \pi/3)$, 
then a standard calculation gives $T_{as}$ and $R_{as}$, and
from them we obtain the value $z_a = -0.44 + 0.49 \ i,$
with the parameters indicated in Fig.~3. Obviously, from these 
(experimental) data alone  we cannot infer at all the possible 
path for this discrete transformation.

However,  the Iwasawa decomposition remedies 
this serious drawback: from the geometrical
meaning discussed before, and once we know the
values of $\phi$, $\xi$, and $\nu$ [that are easily
computed from Eqs.~(\ref{param})] we get, by 
the ordered application of the matrices $\mathsf{K}(\phi)$,
$\mathsf{A}(\xi)$, and $\mathsf{N}(\nu)$,  that the 
trajectory from $z_s$ to $z_a$ is well defined through 
the corresponding orbits, as shown in Fig.~3.

Moreover, and this is the important moral we wish
to extract from this simple example, if in some 
experiment the values of $z_s$ and $z_a$ are
measured, one can find, no matter how complicated 
the multilayer is, in a unique way, the three arcs of orbits 
that connect the initial and final points in the unit disk.

In Fig.~4 we have plotted the values of the
parameters $\phi$, $\xi$, and $\nu$ for this system
when $\beta_2$ is varied between 0 and $\pi$. We have 
also plotted the values of $[\mathrm{Tr}( \mathsf{M}_{as})]^2$. 
It is evident that the system can be of every type 
depending on the value of $\beta_2$. The marked points 
determine special behaviors in agreement with 
Eq.~(\ref{res}) for symmetrical systems: for the left one,  
$\phi=\nu=0$ and the system is  represented by a 
pure matrix $\mathsf{A}(\xi)$;  in the right one, 
$\xi=\nu=0$ and then it is represented by  a matrix  
$\mathsf{K}(\phi)$; i.e., it is antireflection stack
with $T_{as}= \exp(- i \phi/2)$. Note that this system can 
never be represented by a matrix  $\mathsf{N}(\nu)$, 
because it is symmetric.

To show the characteristic properties of an asymmetric system, 
we consider, as indicated in Section 3, the simplest one 
constituted by a two-layer stack made of a glass plate
(with phase thickness $\beta_1$) coated with 
a film  of zinc sulphide (with phase thickness $\beta_2$). In Fig.~5
we have plotted the values of  $\phi$, $\xi$, and $\nu$, as well 
as  $[\mathrm{Tr}( \mathsf{M}_{as})]^2$, when $\beta_1$
is fixed and $\beta_2$ is varied between 0 and $\pi$. From
our previous analysis it is clear that only in the marked point
we have $\phi=\xi=0$ and $[\mathrm{Tr}( \mathsf{M}_{as})]^2= 4$,
so the system is represented by a pure matrix $\mathsf{N}(\nu)$.
Contrary to the previous example, this system can never be 
represented by a matrix  $\mathsf{A}(\xi)$, because it is 
asymmetric.

In summary, we expect that the geometrical scenario
presented here could provide an appropriate tool for 
analyzing and classifying multilayer performance in an
elegant and concise way that, additionally, could be closely
related to other fields of physics.

\bigskip

Luis L. S\'anchez-Soto's e-mail address is
lsanchez@eucmax.sim.ucm.es

\newpage

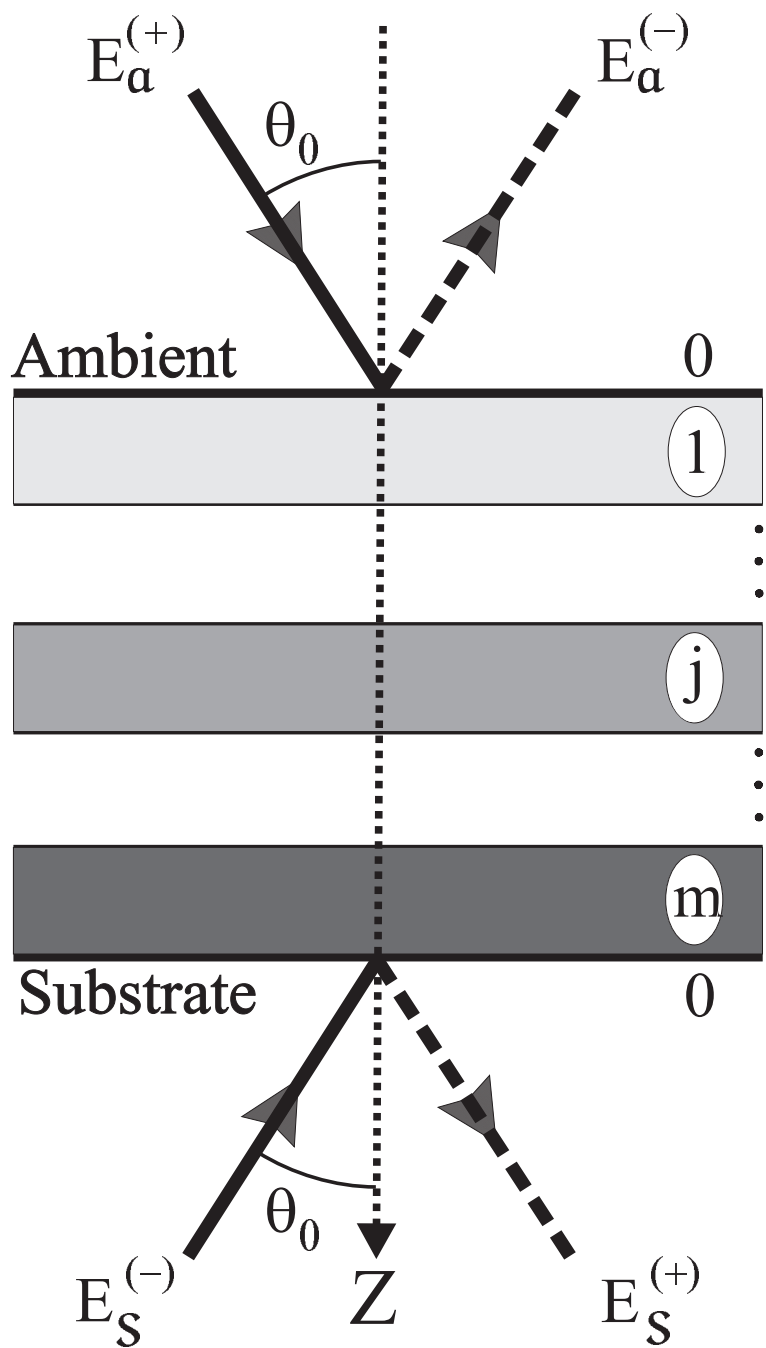
\begin{figure}
\caption{Wave vectors of the input $[E_{a}^{(+)}$ and  
$E_{s}^{(-)}]$ and output $[E_{a}^{(-)}$ and  
$E_{s}^{(+)}]$ fields in a multilayer sandwiched 
between two identical semi-infinite ambient  and substrate 
media.}
\end{figure}

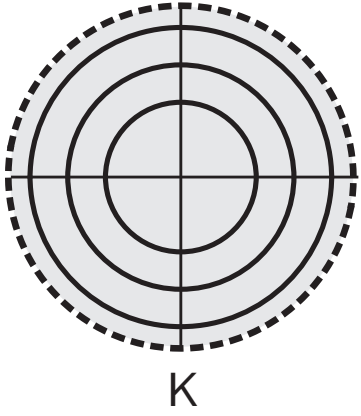
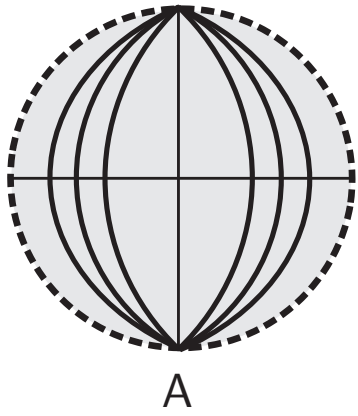
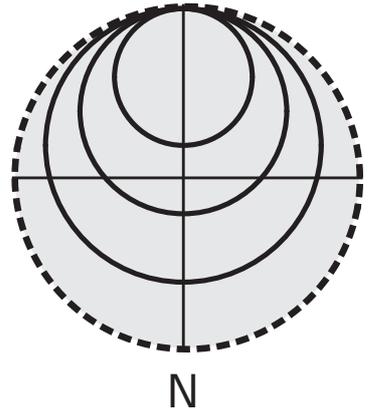
\begin{figure}
\caption{Plot of several orbits in the unit disk 
of the elements of the Iwasawa decomposition 
$\mathsf{K}(\phi)$,  $\mathsf{A}(\xi)$,  and 
$\mathsf{N} (\nu)$ for the group of multilayer
transfer matrices.}
\end{figure}

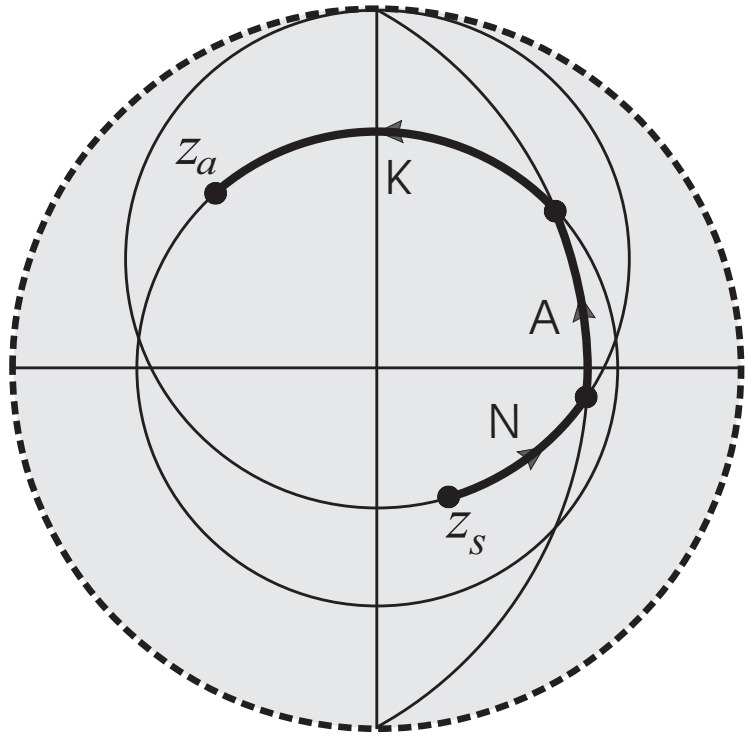
\begin{figure}
\caption{Geometrical representation in the unit disk
of the action of a symmetric system made up of
two  identical plates ($n_1 = 1.7$, $d_1 = 1$~mm,
$\theta_0=\pi/4$, $\lambda = 0.6888~\mu$m and
$s$-polarized light) separated by a spacer of 
phase thickness $\beta_2 =  3$~rad.  The point 
$z_s= 0.4 \exp(- i \pi/3)$ is  transformed by the 
system into the point $z_a= - 0.44 + 0.49 i$. 
We indicate the three orbits given 
by the Iwasawa decomposition and, as a thick line, 
the trajectory associated to the multilayer action.}
\end{figure}

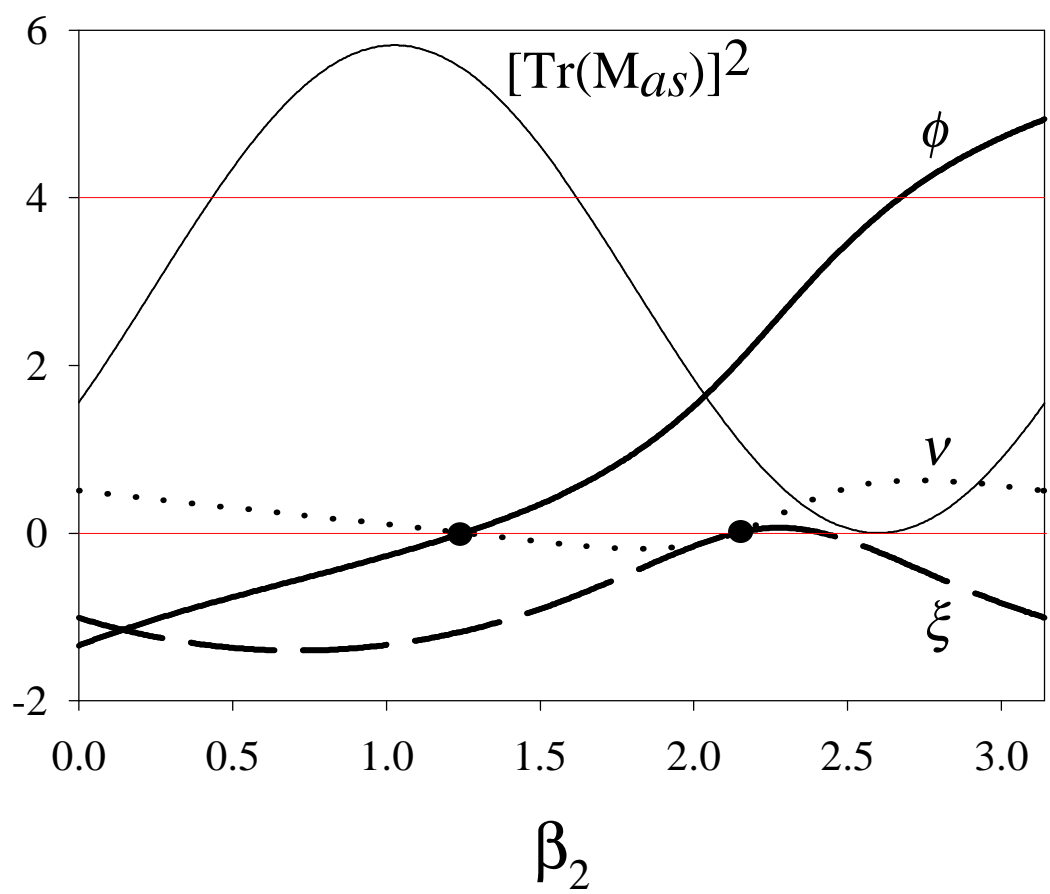
\begin{figure}
\caption{Plot of the values of $[\mathrm{Tr}
(\mathsf{M}_{as})]^2$ and of the parameters
$\phi$, $\xi$, and $\nu$ in the Iwasawa decomposition
for the same system as in Fig.~3, as a function of
$\beta_2$.}
\end{figure}

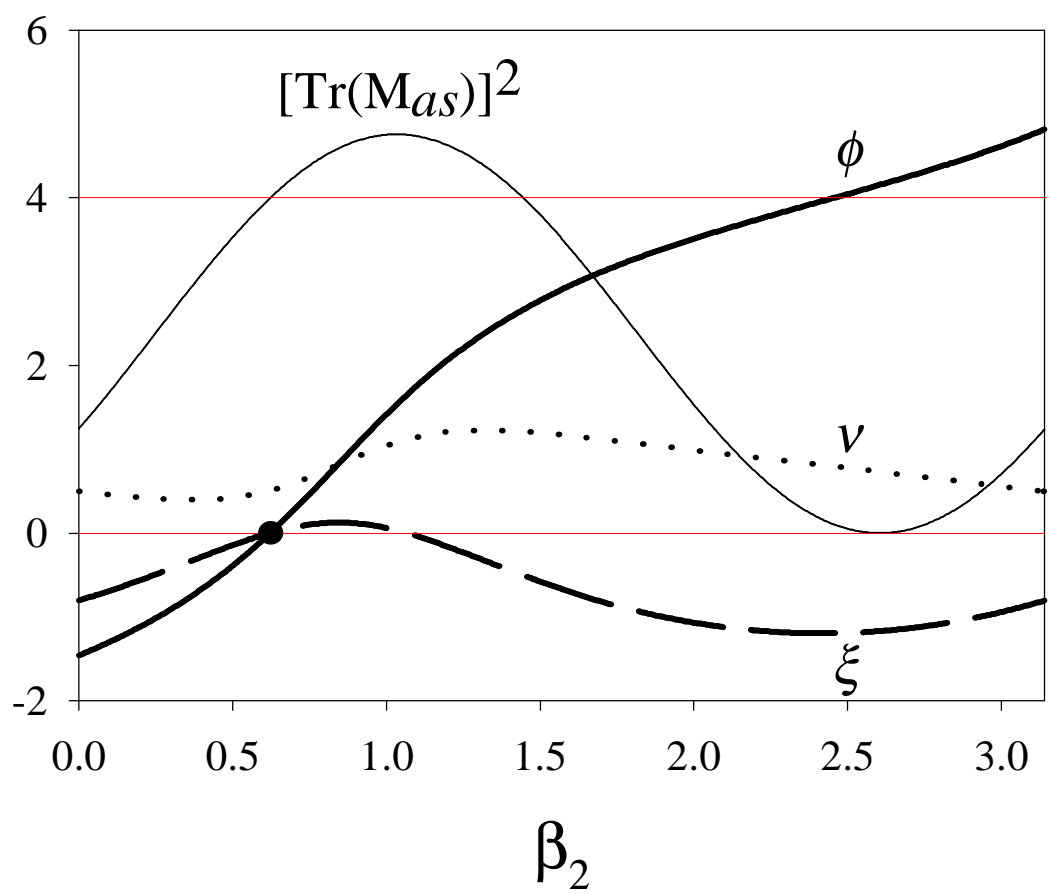
\begin{figure}
\caption{Same plot as in Fig.~4 but for an asymmetric
system made up of a glass plate ($n_1 = 1.5$ and 
$\beta_1 = 2.75$~rad) coated with a zinc sulphide film
($n_2 = 2.3$).}
\end{figure}

\begin{thebibliography}{99}
\bibitem{Snell}
When ambient ($0$) and substrate ($m+1$) media  are 
different,  the angles  $\theta _{0}$ and  $\theta _{m+1}$ 
are connected by Snell law $n_0\sin \theta_0=  
n_{m+1}\sin \theta_{m+1}$, where $n_j$ denotes 
the refractive index of the $j$th medium.
\end{thebibliography}
\end{document}